\documentclass[%
 ,secnumarabic%
,amssymb, amsmath,nobibnotes, aps, prl,
showpacs,showkeys]{revtex4}
\usepackage{docs}%
\usepackage{bm}%
\usepackage{graphicx}
\expandafter\ifx\csname package@font\endcsname\relax\else
 \expandafter\expandafter
 \expandafter\usepackage
 \expandafter\expandafter
 \expandafter{\csname package@font\endcsname}%
\fi

\pacs{25.30.Pt, 13.15.+g} \keywords{neutrino interactions,
 Rein-Sehgal model,  single  pion production, Albright-Jarlskog relations}

\begin{document}

\title{Lepton mass effects in weak CC single pion production}

\author{Krzysztof M. Graczyk}
\email{kgraczyk@ift.uni.wroc.pl}
\author{Jan T. Sobczyk}
\affiliation{ Institute of Theoretical Physics, University of
Wroc\l aw, pl. M. Borna 9, 50-204, Wroc\l aw, Poland}
\date{\today}%

\begin{abstract}
Different approaches to take into account nonzero lepton mass in the
Rein-Sehgal model are compared. Modification of the axial current
due to pion pole term are included and it is shown that they lead to
large reduction of antineutrino cross section and a change of the
shape of $d\sigma/dQ^2$.
\end{abstract}

\maketitle

The analysis of results of neutrino experiments is based on Monte
Carlo generators of events. Because precise experimental data on
inclusive and exclusive cross sections is still missing the
generators rely on approximate models. In the 1 GeV neutrino energy
region there is an important contribution from charged current (CC)
and neutral current (NC) single pion production (SPP) channels. In
most Monte Carlo (MC) codes \cite{MC} the dynamics responsible for
SPP is described in the framework of the Rein-Sehgal (RS) model
\cite{Rein:1980wg}.

The RS model is an old construction based on the relativistic
quark resonance FKR model \cite{FKR}. It is known that in the case
of electro-production its predictions are far away from the new
precise experimental data but nevertheless the model is useful in
describing neutrino interactions. There are several reasons why it
is so. The model is suitable for MC applications as for SPP
channels it provides a description of all degrees of freedom. The
original RS paper provides a clear algorithm how to implement the
model. Another advantage of the RS model is that it covers a large
region in $W$ ($W<2$~GeV) in the kinematically allowed space. But
the most important is that the predictions of the model agree with
the existing data. The predictions for the integrated cross
sections can be fine tuned to the data by modifying the value of
the free parameter, an axial mass. An additional fine tuning can
be done with a non-resonant background which must be added in some
SPP channels. There are other more sophisticated and better
founded approaches to describe SPP in the $\Delta(1232)$ resonance
region like e.g. Lee-Sato model \cite{L-S} but the RS model is
still used and its performance must be regarded as an important
topic of investigation.

The original RS approach adopts the assumption that leptonic mass
$m=0$. With this assumption all the computations in the model become
easier. An approximate way to introduce $m\neq 0$ corrections is to
consider them in the kinematics only. The procedure is to use the
formula for $\frac{d^2\sigma}{dWdQ^2}$ as in the original RS model
but to perform integration over the restricted kinematical region in
the $(W, Q^2)$ plane, the same as in the $m\neq 0$ case. This is the
common way in which $m\neq 0$ effects are included in MC codes.

In recent years there is a growing number of indications that
predictions of MC generators overestimate the cross section in the
low $Q^2$ region. It is therefore important to check exactly the
modifications introduced to the RS model by $m\neq 0$ effects and it
is the subject of this paper. The modifications we consider are of
two kinds. First we use the same hadronic current as in the RS
model. The inclusion of $m\neq 0$ effects requires extra computation
of hadronic weak current matrix elements. Our calculations are based
on the observation that the operational structure of the components
of the hadronic current $\mathcal{J}_0$ and $\mathcal{J}_3$ are
identical. In the original RS paper the matrix elements of linear
combination $\mathcal{J}_{\underline{0}}\equiv \mathcal{J}_0 +
\frac{\nu_{res}}{q_{res}}\mathcal{J}_3$ are calculated. We use these
results and by appropriate substitutions obtain matrix elements of
$\mathcal{J}_0$ and $\mathcal{J}_3$ separately.

\begin{figure}
\centerline{
\includegraphics[width=16cm,height=5cm]{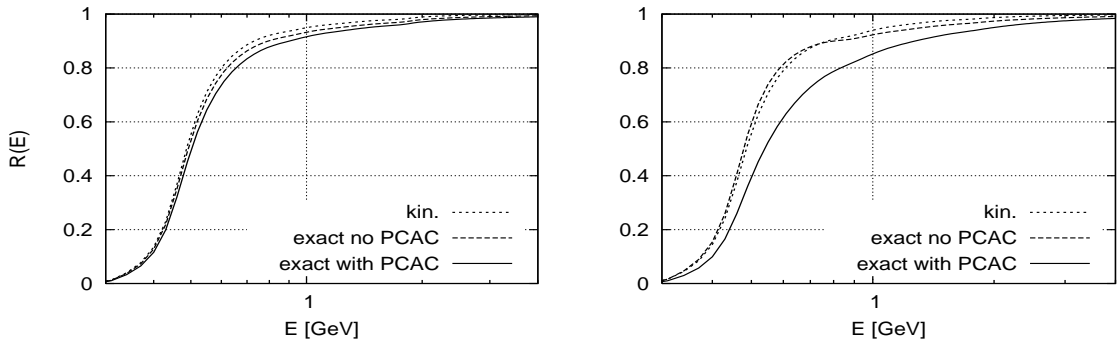}
} \caption{Reduction of the total cross sections for reactions $\nu
n \to \mu^- p \pi^0 $ (left figure) and $\overline{\nu} p \to \mu^+
n \pi^0 $ (right figure) due to $m\neq 0$. Three computations are
compared: (i) kinematical approximation (dotted line) (ii) exact
computation without pion pole contribution (dashed line) (iii) exact
computations with pion pole contribution included (solid line). The
cut on the invariant hadronic mass $W< 2$~GeV is imposed.
\label{FIG_ratio_total_masses}}
\end{figure}

The second modification is more subtle and  requires an inclusion of
a new term in the axial current based on the PCAC arguments. The
procedure how to modify the axial current was described in
\cite{Ravndal73}:

\begin{equation}
\label{axial_current_full_general_form} \mathcal{J}^{A}_\mu
\hookrightarrow \mathcal{J}^{A,mod}_\mu \equiv \mathcal{J}^A_\mu +
q_\mu \frac{q^\nu \mathcal{J}^A_\nu }{m_\pi^2 +Q^2}.
\end{equation}
The pion pole term does not contribute when the lepton mass is
vanishing because the leptonic tensor $L_{\mu\nu}$ satisfies
$L_{\mu\nu}q^{\nu}= -2 m^2 k_\mu $. In the coordinate system in
which $q^\mu=(\nu,0,0,q)$ the PCAC modifies only $\mathcal{J}_0$
and $\mathcal{J}_3$ components of the axial hadronic current and
its inclusion can be again realized by substitutions in the
original RS formulas. Some technical details about our
calculations are contained in
Appendix~\ref{appendix_cross_section}.
\begin{figure}[ht]
\centerline{
\includegraphics[width=16cm,,height=5cm]{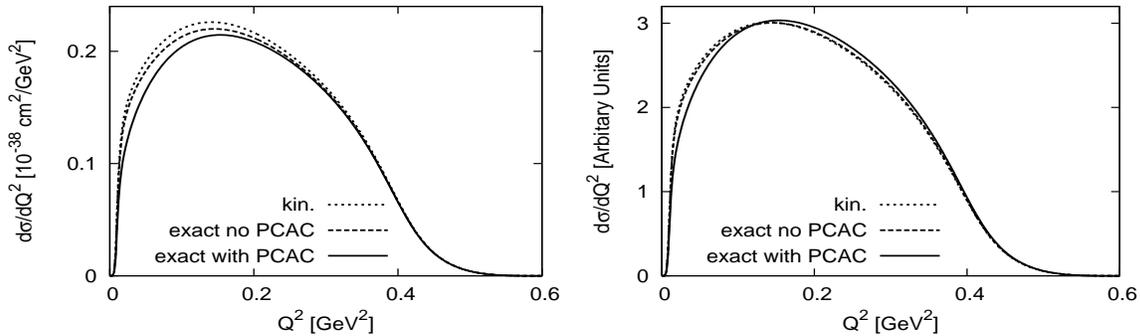}}
\caption{ The differential cross sections for reaction $\nu
n\to\mu^-p\pi^0$ at $E=700$~MeV calculated in three models: (i)
kinematical approximation (dotted line) (ii) exact computation
without pion pole contribution (dashed line) (iii) exact
computations with pion pole contribution included (solid line). In
the right panel the cross sections are normalized to the area under
the curve. The cut on the invariant hadronic mass $W< 2$~GeV is
imposed. \label{FIG_ratio_Q2}}
\end{figure}

\begin{figure}[ht]
\centerline{
\includegraphics[width=16cm,height=5cm]{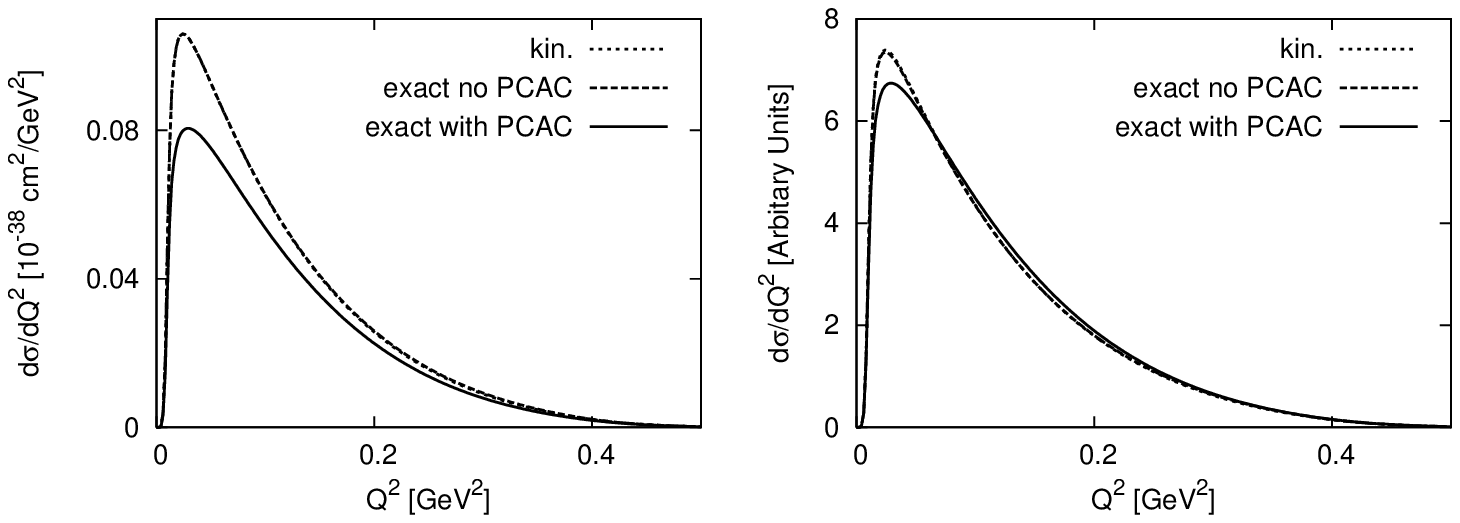}}
\caption{ The same as in Fig. \ref{FIG_ratio_Q2} but for reaction
 $\overline{\nu}p\to\mu^+n\pi^0$.
\label{FIG_ratio_Q2_anty}}
\end{figure}

Few years ago Naumov's group extended the RS model to take into
account the first part of $m\neq 0$ effects \cite{Kuzmin:2004ke}.
They expressed cross section by matrix elements of three currents:
$\mathcal{J}_\pm$ (as in the RS model) and new component
$\widetilde{\mathcal{J}}_{\underline{0}}\neq
\mathcal{J}_{\underline{0}}$. It can be show that this part their
calculations are equivalent to ours. However, there is  an important
difference: in \cite{Kuzmin:2004ke} the pion pole contribution was
not considered and we will see in the discussion that it is quite
relevant.

%%%%%%%%%%%%%%%%%%%%%%%%%%%%%%%%%%%%%%%%%%%%%%%%%%%%%%%%%%%%%%%%%%%%%%%%%%%%%

%%%%%%%%%%%%%%%%%%%%%%%%%%%%%%%%%%%%%%%%%%%%%%%%%%%%%%%%%%%%%%%%%%%%%%%%%%%%%

In Fig.~\ref{FIG_ratio_total_masses} we show the plots with
relative modification of the cross section for $\nu
 n \rightarrow \mu^-p \pi^0$ and $\overline{\nu} p\rightarrow
\mu^+ n \pi^0$ caused by $m\neq 0$.  The functions we define are:

\begin{equation}
\label{ratio_total} \mathcal{R}(E) = \frac{\sigma(E,
m=m_\mu)}{\sigma_{}(E, m= 0)},
\end{equation}
where $\sigma(E, m=m_\mu)$ is calculated in three different model:
\begin{itemize}
\item[(i)]  in {\it kinematical approximation} i.e. including
$m\neq 0$ effects only in kinematics, as described above;

\item[(ii)] in exact computation but without pion pole contribution;

\item[(iii)] in the complete computations with pion pole contribution
included.
\end{itemize}

In the case of neutrino-nucleon scattering all three approaches give
rise to comparable  results. At $E \sim 1$~GeV the differences
between predictions for the total cross section is of order 5 \% .

In the case of antineutrino-nucleon scattering the approaches (i)
and (ii) give similar results while the exact computations (iii)
with pion pole contribution introduce a significant reduction of the
total cross section.

We verified that the same is true also for other neutrino and
antineutrino induced SPP channels.

In the Fig.~\ref{FIG_ratio_Q2} we investigate $d\sigma/dQ^2$ (left
panel) and its shape (right panel) for reaction $\nu
p\to\mu^-p\pi^0$ at neutrino energy $E=0.7$~GeV. Approaches (i) and
(ii) give similar results. The contribution from pion pole visibly
reduces the differential cross section for low $Q^2$ but its shape
remains virtually unchanged. It is an important observation because
in the experimental analysis the shape of $d\sigma/dQ^2$ is of major
interests.

The situation becomes different in the case of antineutrino-nucleon
scattering. In Fig.~\ref{FIG_ratio_Q2_anty} we plot the same curves
as in  Fig.~\ref{FIG_ratio_Q2}  but this time the reaction is
$\overline{\nu}p\to\mu^+n\pi^0$. The peak of the differential cross
section in the approach (iii) is $\sim 25$\% lower in the models (i)
and (ii). What is more important also the shape of the differential
cross section is changed and the peak is reduced by $\sim 10$\% .

\begin{figure}
\centerline{
\includegraphics[width=18cm]{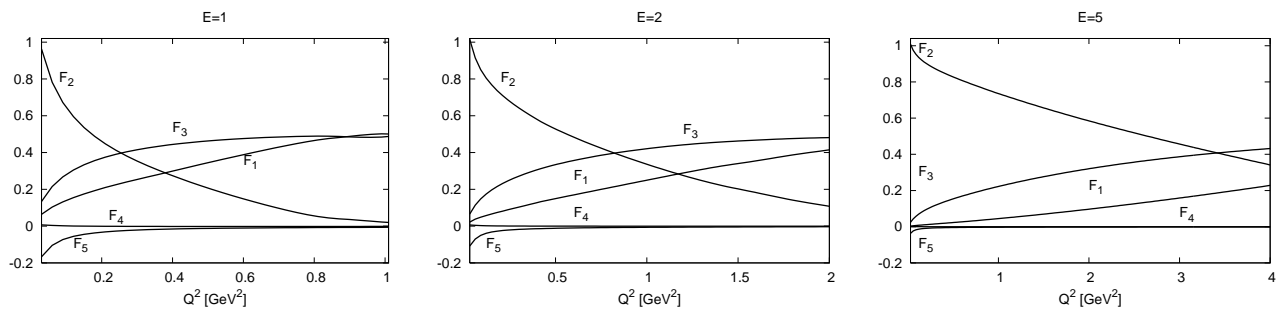}}
\caption{The relative contributions from structure functions $F_j$
in the  differential cross section $d \sigma/dQ^2$ for reaction $\nu
n \to \mu^-p\pi^0$ at neutrino energies 1, 2 and 3~GeV.
\label{Fig_contributions_W}}
\end{figure}

The difference between neutrino and antineutrino scattering can be
understood when the differential cross sections is expressed in
terms of  $F_j$, $j=1,...,5$ structure functions (see Appendix
\ref{Appendix_SF}). Only $F_4$ and $F_5$ become modified by the pion
pole terms but the contribution from the $F_4$ is negligible. The
contribution from $F_5$ is negative (pion pole terms make it
smaller) and its absolute values decreases quickly with $Q^2$ (see
Fig. \ref{Fig_contributions_W}). Therefore, it reduces the
differential cross section in low $Q^2$. In the case of antineutrino
scattering the $F_3$ terms contributes with the negative sign. The
antineutrino cross section is smaller and the $F_5$ induced
reduction is more important.

There is another approximate way to include $m\neq 0$ effects: to
apply exact expressions for $F_{1,2,3}$ as defined in the original
RS model and to use Albright-Jarlskog relations (A-J)
\cite{Albright:1974ts} for $F_4$ and $F_5$:
\begin{equation}
\label{A-J_relation} F_4=0, \qquad xF_5=F_2
\end{equation}

We checked numerically that $xF_5=F_2$ holds with an accuracy of
20\% for $Q^2\sim 0.05$~GeV$^2$,  $10\%$ for $Q^2\sim 0.1$~GeV$^2$
and 1\% for $Q^2\sim 0.5$~GeV$^2$.  The ratio $xF_5/F_2(W)$ depends
very weakly on the hadronic invariant mass.

\begin{figure}
\centerline{
\includegraphics[width=16cm,height=5cm]{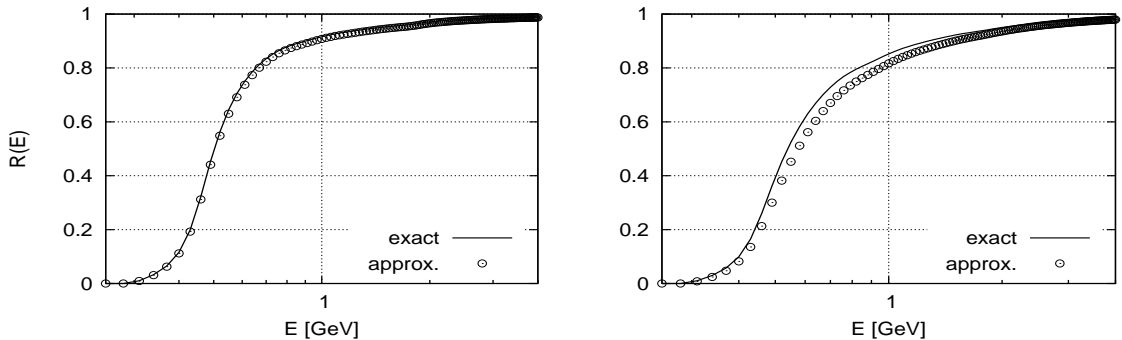}
} \caption{ Reduction of the total cross sections for reactions $\nu
n \to \mu^- p \pi^0 $ (left figure) and $\overline{\nu} p \to \mu^+
n \pi^0 $ (right figure) due to $m\neq 0$. Two computations are
compared: exact computations with pion pole contribution included
(solid line) and the approximation based on Albright-Jarlskog
relations (circles). The cut on the invariant hadronic mass $W<
2$~GeV is imposed. \label{FIG_ratio_total_RS_vs_FF}}
\end{figure}
\begin{figure}
\centerline{
\includegraphics[width=16cm,height=5cm]{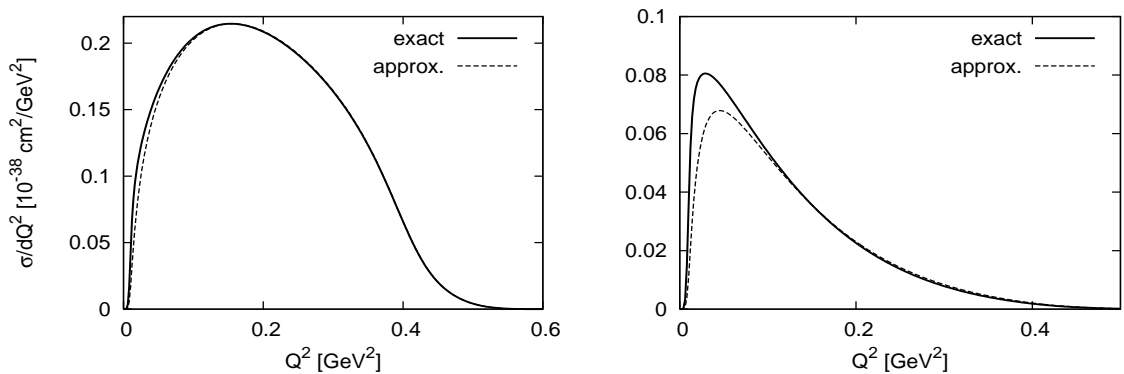}
} \caption{Differential cross section for reactions $\nu n\to
\mu^-\pi^0p $ (left figure) and $\overline{\nu} p\to \mu^+\pi^0n $
(right figure) for E=700 MeV. Cross sections are obtained in exact
computation (solid lines) and in approximation with  $F_4$ and $F_5$
calculated with Albright-Jarlskog relations (dashed
line).\label{FIG_diff_Q2_RS_vs_FF}}
\end{figure}

In Fig.~\ref{FIG_ratio_total_RS_vs_FF} we present the similar plots
as in Fig.~\ref{FIG_ratio_total_masses}. We see that for neutrino
scattering the approximation based on the A-J relation is a good one
but in antineutrino case the total cross section is underestimated.
In Fig. \ref{FIG_diff_Q2_RS_vs_FF} we see that at $E=700$~MeV the
approximation we discuss reconstructs well the $d\sigma/dQ^2$ in the
case of neutrino reaction. For antineutrino scattering the
approximation based on the A-J relations fails to reproduce both
$d\sigma/dQ^2$ and its shape.

The conclusions of our investigation is that (in the case of
antineutrino reactions) it is improper to use RS model without pion
pole terms. The approximation based on the A-J relation is
satisfactory only for neutrino scattering and not in the
antineutrino case.

After this paper was completed we learned about the work of Ch.
Berger and L. Sehgal \cite{Sehgal} in which the same problem is
discussed. The authors of Ref. \cite{Sehgal} use the formalism
developed in \cite{Kuzmin:2004ke} and in their presentation focus on
modifications of the cross sections at small scattering angles.

%%%%%%%%%%%%%%%%%%%%%%%%%%%%%%%%%%%%%%%%%%%%%%%%%%%%%%%%%%%%%%%%%%%%%%%
%%%%%%%%%%%%%%%%%%%%%%%%%%%%%%%%%%%%%%%%%%%%%%%%%%%%%%%%%%%%%%%%%%%%%%%
\appendix
%%%%%%%%%%%%%%%%%%%%%%%%%%%%%%%%%%%%%%%%%%%%%%%%%%%%%%%%%%%%%%%%%%%%%%%
\section{}
\label{appendix_cross_section}  Incoming neutrino and outgoing
charged lepton 4-momenta are denoted as $k^\mu$ and ${k'}^\mu$.
Similarly, $p^\mu$ and ${p'}^\mu$ are target nucleon and outgoing
resonance 4-momenta. 4-momentum transfer is $q^\mu \equiv k^\mu -
{k'}^\mu = {p'}^\mu - p^\mu$, $Q^2\equiv - q_\mu q^\mu$, $k^\mu =
(E, \mathbf{k})$. In the Lab frame the axis orientation is chosen so
that $q^\mu = (\nu, 0, 0, q)$. $M$ denotes the nucleon's and $M_{R}$
the resonance mass, $W$ is the invariant hadronic mass of the final
state, charged lepton mass is denoted by $m$. Explicit computations
are done in the resonance rest frame and then the 4-vectors are
labeled by subscripts $_{res}$.

In the RS model the differential cross section for
neutrino/antineutrino- resonance production are expressed by
matrix elements of $ \mathcal{J}_0$, $\mathcal{J}_3$,
$\mathcal{J}_+\equiv -\frac{1}{\sqrt{2}}\left(\mathcal{J}_1 + i
\mathcal{J}_2 \right)$, $\mathcal{J}_-\equiv
\frac{1}{\sqrt{2}}\left(\mathcal{J}_1 -i \mathcal{J}_2 \right)$:

\begin{eqnarray}
\label{diff_cross_section_nuN} {\frac{d^2 \sigma}{d \nu d Q^2}} &=&
\frac{ G^2 \cos^2\theta_C}{ 4 \pi E^2} (2\pi)^6
\overline{\sum_{s,s'}} \frac{E_{p,res}}{M} \delta(W - M_{R})
\times \nonumber \\
& &\left\{ D_0 \left|\left<{p'}_{res},s'\right| \mathcal{J}_0
\left|p_{res},s\right>\right|^2  +  D_3
\left|\left<{p'}_{res},s'\right| \mathcal{J}_3
\left|p_{res},s\right>\right|^2  -2 D_{03}
\mathrm{Re}\left(\left<{p'}_{res},s'\right| \mathcal{J}_0
\left|p_{res},s\right> \left<{p}_{res},s\right|
\mathcal{J}_3 \left|{p'}_{res},s'\right>\right) \right. \nonumber \\
& & \left.  +(A \mp B) \left|\left<{p'}_{res},s'\right|
\mathcal{J}_+ \left|p_{res},s\right>\right|^2 + (A\pm
B)\left|\left<{p'}_{res},s'\right| \mathcal{J}_-
\left|p_{res},s\right>\right|^2\right\}
\end{eqnarray}
where $\mp$ refers to neutrino/antineutrino.
\begin{eqnarray*}
D_0    = 2 E^{2}_{res} - 2\nu_{res} E_{res}+ k_\mu q^\mu, \quad D_3
&=& 2 (k^3_{res})^{2} - 2 q_{res} k^{3}_{res} - k_\mu q^\mu,\quad
D_{03} = \nu_{res} k^{3}_{res} +q_{res} E_{res} - 2E_{res}
k^{3}_{res} \\
A &=& |k_+^{res}|^2 - k_\mu q^\mu, \quad B = \left(  q_{res}
E_{res}-\nu_{res} k^{3}_{res}\right)
\end{eqnarray*}
and also
\begin{equation}
k^{3}_{res}            =  \frac{m^2 +Q^2
+2E_{res}\nu_{res}}{2q_{res}}, \quad |k_+^{res}|^2 = E^{2}_{res} -
(k^{3}_{res})^2. %\quad k_\mu q^\mu =-\frac{m^2 + Q^2}{2}
\end{equation}
In the limit $m\rightarrow 0$ the cross section
(\ref{diff_cross_section_nuN}) is expressed in terms of matrix
elements of $\mathcal{J}_-$, $\mathcal{J}_+$ and
$\mathcal{J}_{\underline{0}}\equiv \mathcal{J}_0 +
\frac{\nu_{res}}{q_{res}}\mathcal{J}_3$. Their values for 18
resonances are listed in Tab. II of Ref.~\cite{Rein:1980wg}. In our
computation we use the normalization from our previous paper
\cite{GS_form_factors}. In practice, functions $S$, $B$ and $C$ of
this paper differ by the factor $1/2W$ with respect to analogical
functions introduced in \cite{Rein:1980wg}. ($S_{{this}\;{paper}}= 2
W S_{RS}$,... etc.).

Matrix elements of $\mathcal{J}_0$ and $\mathcal{J}_3$ can be
obtained because in the RS model they have the same operational
structure as $\mathcal{J}_{\underline{0}}$:

\begin{equation}
\mathcal{J}_{\underline{0}}^V(S)= 9\tau_a^+ S e^{-\lambda
a^{3\dagger}},\quad \mathcal{J}_{\underline{0}}^A(B,C)= - 9 \tau^+_a
e^{-\lambda a^{3\dagger}}\left(C \sigma^3_a + B\vec{\sigma}_a\cdot
\vec{a}^\dagger \right),\quad  S= \frac{Q^2}{q_{res}^2}\frac{3WM
-Q^2 -M^2}{3W^2}G_V
\end{equation}
and all we need to do is to make substitutions:
\begin{equation}
\mathcal{J}_0^V  =  \mathcal{J}^V_{\underline{0}}\left( S \to S_0=
\frac{q^{2}_{res}}{Q^2}S\right), \quad \mathcal{J}_3^V  =
\mathcal{J}^V_{\underline{0}}\left( S \to S_3=
-\frac{\nu_{res}q_{res}}{Q^2}S\right)
\end{equation}
and
\begin{equation}
\mathcal{J}_0^A  =  \mathcal{J}^A_{\underline{0}}\left( B\to
B_0,C\to C_0  \right), \quad \mathcal{J}_3^A  =
\mathcal{J}^A_{\underline{0}}\left(B\to B_3, C\to C_3\right) .
\end{equation}

with

\begin{eqnarray}
B_0 &=&   G_A Z\frac{2}{3}\sqrt{\frac{\Omega}{2}},   \quad C_0 =
G_A Z \frac{M q}{W}
\left(\frac{1}{3} + \frac{ W^2 -Q^2 -M^2}{(W+M)^2 +Q^2}\right),\\
B_3 &=&   G_A  Z \frac{4 M q}{3\left((W+M)^2
+Q^2\right)}\sqrt{\frac{\Omega}{2}},\\
C_3 &=&   G_A Z  \left( \frac{3W^2 +Q^2 +M^2}{6} - \frac{2W}{(W+M)^2
+Q^2}\left( q^{2}\frac{M^2}{W^2} + \frac{N}{3} \Omega\right),
 \right)
\end{eqnarray}
where $\lambda = \sqrt{\frac{2}{\Omega}}q_{res}$, $Z=0.7602$,
$\Omega =$1.05~GeV$^2$ is determined from the Regge slope of baryon
trajectories.

In order to include pion pole terms we calculate:
\begin{equation}
q^{\mu}_{res} \mathcal{J}^A_\mu =\mathcal{J}^A_{\underline{0}}\left(
B\to B_D\equiv \nu_{res}B_0 + q_{res}B_3,\quad C\to C_D\equiv
\nu_{res}C_0+q_{res}C_3.\right).
\end{equation}
and the final
expressions for the axial current are:
\begin{eqnarray}
\mathcal{J}^{A,mod}_0 &=&\mathcal{J}_{\underline{0}}\left( B\to B_0+
\nu_{res}\frac{B_D}{m_\pi^2+Q^2},\quad C\to C_0+
\nu_{res}\frac{C_D}{m_\pi^2+Q^2}\right),\\
\mathcal{J}^{A,mod}_3 &=&\mathcal{J}_{\underline{0}}\left( B\to
B_3- q_{res}\frac{B_D}{m_\pi^2+Q^2},\quad C\to C_3-
q_{res}\frac{C_D}{m_\pi^2+Q^2}\right).
\end{eqnarray}

%%%%%%%%%%%%%%%%%%%%%%%%%%%%%%%%%%%%%%%%%%%%%%%%%%%%%%%%%%%%%%%%%%%%%%%

\section{Structure functions}
\label{Appendix_SF}

The cross section for neutrino/antineutrino-nucleon scattering has
the form:
\begin{eqnarray}
\label{cross_section_Wi} {\frac{d^2 \sigma}{d \nu d Q^2}} &= &
\frac{ G^2 \cos^2\theta_C}{ 4\pi E^2} \left\{
(Q^2+m^2)\frac{F_1}{M} + \left( 2E(E-\nu )-
\frac{m^2+Q^2}{2}\right) \frac{F_2}{\nu}\right. \nonumber\\
& & \left. \pm \left( EQ^2-\frac{\nu}{2}(m^2+Q^2)
\right)\frac{F_3}{\nu M} + \frac{m^2}{2}\left( Q^2 +
m^2\right)\frac{F_4}{\nu M^2}-\frac{m^2E}{\nu M}F_5 \right\} .
\end{eqnarray}
where
\begin{eqnarray} F_1 & = & (2\pi)^6 \delta(W - M_{R})
\frac{M}{2} \overline{\sum_{s,s'}}
\left\{\left|\left<p'_{res},s'\right| \mathcal{J}_-
\left|p_{res},s\right>\right|^2 + \left|\left<p'_{res},s'\right|
\mathcal{J}_+ \left|p_{res},s\right>\right|^2
\right\}\frac{E_{p,res}}{M}, \\
F_2 & = & (2\pi)^6 \delta(W - M_{R}) \frac{E_{p,res}}{M}\frac{\nu
Q^2}{2
q^{2}}\nonumber\\
& &\!\!\!\!\!\overline{\sum_{s,s'}}\ \left\{
\frac{2q_{res}^{2}}{Q^2}\left|\left<p'_{res},s'\right|
\mathcal{J}_{\underline{0}} \left|p_{res},s\right>\right|^2
 +  \left|\left<p'_{res},s'\right|\mathcal{J}_-
\left|p_{res},s\right>\right|^2 + \left|\left<p'_{res},s'\right|
\mathcal{J}_+
\left|p_{res},s\right>\right|^2  \right\}, \\
F_3 &=& (2\pi)^6 \frac{\nu M}{q}\delta(W - M_{R})
\overline{\sum_{s,s'}} \left\{\left|\left<p'_{res},s'\right|
\mathcal{J}_- \left|p_{res},s\right>\right|^2 -
\left|\left<p'_{res},s'\right| \mathcal{J}_+
\left|p_{res},s\right>\right|^2 \right\}\frac{E_{p,res}}{M},
\end{eqnarray}
\begin{eqnarray}
F_4 & = & (2\pi)^6 \delta(W - M_{R}) \frac{E_{p,res}}{M}
\frac{\nu}{ q^{2}}\overline{\sum_{s,s'}}\left[
\left|\left<p'_{res},s'\right| q_{res} \mathcal{J}_{\underline{0}}
- W  \mathcal{J}_3
\left|p_{res},s\right>\right|^2 \right. \nonumber   \\
& & \left. - \frac{M^2}{2} \left(\left|\left<p'_{res},s'\right|
\mathcal{J}_-  \left|p_{res},s\right>\right|^2 +
\left|\left<p'_{res},s'\right|\mathcal{J}_+
\left|p_{res},s\right>\right|^2\right)\right],  \\
F_5 & = &  (2\pi)^6 \delta(W - M_{R})
\frac{E_{p,res}}{M}\frac{\nu}{2 q^{2}}
\overline{\sum_{s,s'}}\left[ \left|\left<p'_{res},s'\right| 2
q_{res} \mathcal{J}_{\underline{0}} -
W\mathcal{J}_3\left|p_{res},s\right>\right|^2
-W^2\left|\left<p'_{res},s'\right| \mathcal{J}_3
\left|p_{res},s\right>\right|^2 \right.  \nonumber \\
& & \left. - \left(M^2 -Q^2 -
W^2\right)\left(\left|\left<p'_{res},s'\right|
\mathcal{J}_-\left|p_{res},s\right>\right|^2 +
\left|\left<p'_{res},s'\right|\mathcal{J}_+
\left|p_{res},s\right>\right|^2\right) \right].
\end{eqnarray}

In order to calculate $F_1$, $F_2$ and $F_3$ it is enough to know
matrix elements of $\mathcal{J}_\pm$ and
$\mathcal{J}_{\underline{0}}\equiv
\mathcal{J}_0+\frac{\nu_{res}}{q_{res}}\mathcal{J}_3$ and they are
provided in the original RS model. Calculation of $F_4$, $F_5$
requires an additional knowledge of matrix elements of
$\mathcal{J}_3$. In the limit $m\rightarrow 0$ the contribution from
$F_4$ and $F_5$ to cross section vanishes.

$F_1$, $F_2$ and $F_3$ do not depend on pion pole terms because they
do not modify $\mathcal{J}_\pm$ and $\mathcal{J}_{\underline{0}}$.

\section*{Acknowledgements}
\begin{acknowledgments}
The authors were supported by the Polish Ministery of Science grant
 4983/PB/IFT/06.
\end{acknowledgments}

\end{document}